\documentclass[aps,     groupedaddress, showkeys, showpacs]{revtex4}

\usepackage[small]{caption}
\usepackage{amsmath,amsfonts,amscd,amssymb,epsf, epsfig}\usepackage{graphicx}

\newcommand{\Z}{{\mathbb{Z}}}

\newcommand{\R}{\mathbb{R}}

\newcommand{\pa}{\partial}

\begin{document}

\title{Finite temperature Casimir effect in spacetime with extra compactified dimensions}

\author{L.P. Teo}\email{lpteo@mmu.edu.my}\affiliation{Faculty of Information
Technology, Multimedia University, Jalan Multimedia, Cyberjaya,
63100, Selangor Darul Ehsan, Malaysia.}

\keywords{Casimir effect, finite temperature, extra compactified dimensions, Dirichlet boundary conditions.}
\pacs{11.10.Wx, 11.10.Kk, 04.62.+v}

\begin{abstract}
In this letter, we derive the explicit exact formulas for the finite temperature Casimir force acting on a pair of parallel plates in the presence of extra compactified dimensions within the framework of Kaluza-Klein theory. Using the piston analysis, we show that at any temperature, the Casimir force is always attractive and the effect of extra dimensions becomes stronger when the size or number of the extra dimensions increases. These properties are not affected by the explicit geometry and topology of the Kaluza-Klein space.
\end{abstract}
\maketitle

Theories with extra space dimensions can be dated back to the work of Kaluza and Klein \cite{1,2} in an attempt to unify gravity and classical electrodynamics.  The advent of string theory has made the idea of extra   dimensions indispensable \cite{15}. Recently, the interest for physical models that contain extra warped dimensions come from the endeavor  to explain the large gap between the Planck and electroweak scale (the hierarchy problem), and the dark energy and  cosmological constant problem \cite{3, 4, 5, 6, 7, 8, 9, 10, 11, 12, 28, 32, 65, 67, 33, 30, 66, 29, 31, 13, 14, 39}. In this letter, we are interested in studying the Casimir effect in the presence of extra compactified dimensions. The research on Casimir effect in the scenarios with extra dimensions have been studied in the context of string theory \cite{16, 17, 18, 19}, dark energy and cosmological constant \cite{28, 32, 33, 30, 29, 31, 39, 20, 49, 50, 51, 53, 21, 22, 47},  as well as   stabilization of extra dimensions \cite{13, 14, 73, 74, 48, 23, 52, 55, 41, 42, 43, 24, 25, 44, 45, 26, 27, 54}.
 Casimir force for massless scalar field with Dirichlet boundary conditions on a pair of parallel plates inside space with extra dimensions compactified to a torus $T^n = (S^1)^n$   have been calculated in \cite{34, 35, 36, 68, 37, 38}. The dependence of the Casimir energy     on a cut-off scale was investigated in \cite{39}. Casimir effect for electromagnetic fields confined between a pair of parallel and  perfectly conducting plates in $\R^{d+1}$ was considered in \cite{40}. In the case the extra dimension is compactified to a circle $S^1$, it was studied in \cite{46, 39, 56, 57}. For the Randall-Sundrum spacetime model, the Casimir force due to massless scalar fields subject to Dirichlet boundary conditions on two parallel plates were calculated in \cite{58, 59, 60}.

Despite the importance of the thermal corrections, most of the works mentioned above dealt with Casimir effect at zero temperature. In particular, we would like to mention the two papers by  Cheng \cite{36, 37} where he calculated the Casimir force acting on a pair of parallel plates due to massless scalar field with Dirichlet boundary conditions in   spacetime with $n$ extra dimensions   compactified to a torus. As pointed out in the comment \cite{38}, the earlier paper \cite{36} concluded that the presence of extra dimensions will give rise to repulsive Casimir force under certain conditions. However, this result was invalidated by the author himself in the later paper \cite{37}, where he redid the calculations in a more physical setup known as piston \cite{61} and concluded that the Casimir force shall always be attractive.   The finite temperature Casimir energy in $(d+1)$-dimensional rectangular cavities were first calculated in \cite{62} and revisited in \cite{63}.  In \cite{64}, we calculated the finite temperature Casimir force on a $(d+1)$-dimensional piston inside a rectangular cavity due to massless scalar field as well as electromagnetic field and concluded that the Casimir force is always attractive when one end of the cavity was opened.   As a result, it will be natural to suspect the validity of the result of   Cheng \cite{35}, who concluded the possible repulsive   Casimir force  at finite temperature for a pair of parallel plates in space with an extra dimension compactified to a circle.

\begin{figure}\centering \epsfxsize=.5\linewidth
\epsffile{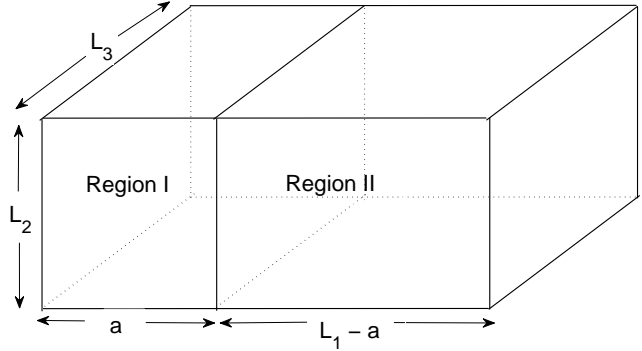}\caption{\label{f1} A three dimensional
rectangular piston}\end{figure}

In this letter, we consider the Casimir force at finite temperature within the piston setup in spacetime with $n$ extra dimensions compactified to a torus $T^n = (S^1)^n$. More precisely, we consider a piston inside a three dimensional rectangular cavity (see FIG. \ref{f1}) in a space with $n$-extra dimensions and calculated the Casimir force acting on the piston for a massless scalar field with Dirichlet boundary conditions on the walls of the rectangular cavity and the piston. At the end, we let $L_1, L_2, L_3$ approach infinity so that we obtain finite temperature Casimir force acting on parallel plates in the presence of extra compactified dimensions. We are also going to comment on the case where the extra dimensions are compactified to an arbitrary compact manifold without boundary.

For a massless scalar field   in the spacetime $M^4\times T^n$, where $M^4$ is the $(3+1)$-dimensional Minkowski spacetime, a complete set of eigenfrequencies of the field confined within a rectangular cavity $[0, L_1]\times[0, L_2]\times[0, L_3]$ subject to the Dirichlet boundary conditions are given by
\begin{equation*}\begin{split}
&\omega_{\boldsymbol{k}, \boldsymbol{l}} = c\sqrt{\left(\frac{\pi k_1}{L_1}\right)^2+\left(\frac{\pi k_2}{L_2}\right)^2+\left(\frac{\pi k_3}{L_3}\right)^2+\sum_{j=1}^n \left(\frac{l_j}{R_j}\right)^2},\end{split}
\end{equation*}where $c$ is the light speed, $\boldsymbol{k}=(k_1, k_2, k_3)\in  \mathbb{N}^3$ and $\boldsymbol{l}=(l_1, \ldots, l_n) \in \Z^n$. Notice that in contrast to \cite{34, 35, 36, 68, 37}, our $l_j, j=1, \ldots, n$ runs from $-\infty$ to $\infty$ instead of $0$ to $\infty$. As pointed out in \cite{38}, this is more natural since the extra dimensions are compactified to a torus, a manifold without boundary. Therefore for each compact dimension, we need to consider a complete set of eigenvalues for Laplace operators on $S^1$ counting with multiplicities.

The finite temperature Casimir energy due to the vacuum fluctuation of the field inside the rectangular cavity is defined by the mode sum:
\begin{equation*}\begin{split}
 E_{\text{Cas}}^{\text{cavity}}(  T) =&\frac{\hbar}{2}\sum \omega + k_BT\sum \log\left(1- e^{-\hbar \omega/ k_B T}\right),\end{split}
\end{equation*}where the summations run through all eigenfrequencies $\omega$, $T$ is the temperature at equilibrium, $\hbar$ is the reduced Planck constant and $k_B$ is the Boltzman constant. For regularization purpose, we introduce a  cut-off $\lambda$ so that
\begin{equation*}\begin{split}
E_{\text{Cas}}^{\text{cavity}}(\lambda;   T) =&\frac{\hbar}{2}\sum \omega e^{-\lambda \omega} + k_BT\sum \log\left(1- e^{-\hbar \omega/ k_B T}\right).\end{split}
\end{equation*}
As pointed out in \cite{61}, the Casimir energy of the piston system (FIG. \ref{f1}) is the sum
\begin{equation*}\begin{split}
E_{\text{Cas}}(\lambda; a;   T)=& \left.E_{\text{Cas}}^{\text{cavity}}(\lambda;   T)\right|_{L_1\rightarrow a}+\left.E_{\text{Cas}}^{\text{cavity}}(\lambda;  T)\right|_{L_1\rightarrow L_1-a}+E_{\text{Cas}}^{\text{ext}}(T)\end{split}
\end{equation*}of the Casimir energies of Region I, Region II and the exterior region. Being independent of the position of the piston, the Casimir energy of the exterior region do not contribute to the Casimir force acting on the piston. As a result,  the Casimir force acting on the piston is given by
\begin{equation*}\begin{split}
F_{\text{Cas}}(a;  T) = -\lim_{\lambda\rightarrow 0^+}&\frac{\pa }{\pa a} \left\{\left.E_{\text{Cas}}^{\text{cavity}}(\lambda;   T)\right|_{L_1\rightarrow a} +\left.E_{\text{Cas}}^{\text{cavity}}(\lambda;  T)\right|_{L_1\rightarrow L_1-a}\right\}.
\end{split}\end{equation*}A distinction about the piston setup is that the Casimir force $F_{\text{Cas}}(a;  T)$ is finite in the $\lambda\rightarrow 0^+$ limit.
Using the same method of computation as presented in \cite{64}, we find that in the  $L_1  \rightarrow \infty$ limit,
\begin{equation}\label{eq12_18_1}\begin{split}
F_{\text{Cas}}(a; T) = -\pi k_BT \sum_{(k_2, k_3)\in\mathbb{N}^2}\sum_{\boldsymbol{l}\in \Z^n} \sum_{p=-\infty}^{\infty}\frac{\Lambda_{k_2, k_3, \boldsymbol{l}, p} }{ \exp\Bigl(2\pi  a \Lambda_{k_2, k_3, \boldsymbol{l}, p}\Bigr)-1} \end{split}
\end{equation} where\begin{equation*}\begin{split}\Lambda_{k_2, k_3, \boldsymbol{l}, p}=\sqrt{\left(\frac{k_2}{L_2}\right)^2+\left(\frac{k_3}{L_3}\right)^2+\frac{1}{\pi^2}\sum_{j=1}^n \left( \frac{l_j}{R_j}\right)^2+\left(\frac{2p k_BT}{\hbar c}\right)^2}.
\end{split}\end{equation*}Eq. \eqref{eq12_18_1} shows clearly that for an opened piston, the Casimir force is always attractive at \emph{any temperature } and the magnitude of the Casimir force is a decreasing function of the distance $a$ between the piston and the opposite wall. Moreover, since each individual term in the summation of \eqref{eq12_18_1} is positive, it shows that the magnitude of the Casimir force becomes larger as the number of extra dimensions increases. In addition, since the function $x\mapsto [x/(e^x-1)], \;x>0$ is a decreasing function, we can infer from \eqref{eq12_18_1} that the Casimir force becomes stronger as the size of the extra dimensions increases. On the other hand,
\eqref{eq12_18_1} also shows that as the size of the extra dimensions shrinks to zero, i.e. $R_j\rightarrow 0$ for $ j=1,\ldots, n$, we have
\begin{equation}\label{eq12_19_1}\begin{split}
\lim_{\substack{R_j\rightarrow 0\\ j=1, \ldots, n}}F_{\text{Cas}}(a; T)=F_{\text{Cas}}^{3D}(a; T) =-\pi k_BT \sum_{(k_2, k_3)\in\mathbb{N}^2} \sum_{p=-\infty}^{\infty}\frac{\sqrt{\left(\frac{k_2}{L_2}\right)^2+\left(\frac{k_3}{L_3}\right)^2 +\left(\frac{2p k_BT}{\hbar c}\right)^2}}{ \exp\left(2\pi  a \sqrt{\left(\frac{k_2}{L_2}\right)^2+\left(\frac{k_3}{L_3}\right)^2 +\left(\frac{2p k_BT}{\hbar c}\right)^2}\right)-1},\end{split}
\end{equation}which corresponds to the $\boldsymbol{l}=\mathbf{0}$ term in \eqref{eq12_18_1}. This implies  that in the limit of vanishing extra dimensions, the finite   temperature Casimir force reduces to the corresponding Casimir force acting on an opened piston for massless scalar field with Dirichlet boundary conditions in (3+1)-dimensional spacetime. Moreover, the correction term of the Casimir force due to the extra dimensions   increases the magnitude of the Casimir force, and it goes to zero exponentially fast as the size of the extra dimensions goes to zero.

To obtain the Casimir pressure
on a pair of infinite parallel plates, we divide the Casimir force \eqref{eq12_18_1} by $L_2L_3$ and take the limits $L_2, L_3\rightarrow \infty$, i.e.,\begin{equation*}
P_{\text{Cas}}^{\parallel}(a;T)=\lim_{L_2, L_3\rightarrow \infty} \frac{F_{\text{Cas}}(a;T)}{L_2L_3}.
\end{equation*} The  detail calculations is similar to those given in \cite{64}. The results can be written as a sum of the Casimir pressure for a pair of infinite parallel plates in (3+1)-dimensional spacetime and the correction term:
\begin{equation}\label{eq12_22_5}
P_{\text{Cas}}^{\parallel}(a;T)=P_{\text{Cas}}^{\parallel, 3D}(a;T) + P_{\text{Cas}}^{\parallel, C}(a;T),
\end{equation}where the the Casimir pressure for a pair of infinite parallel plates in (3+1)-dimensional spacetime is  \begin{widetext}
\begin{equation}\label{eq12_22_1}\begin{split}
&P_{\text{Cas}}^{\parallel, 3D}(a;T)=-\frac{\pi^2 \hbar c  }{480 a^4 }-\frac{\pi^2(k_BT)^4}{90(\hbar c)^3}  +\frac{\pi k_B T}{2a^3}\sum_{k_1=1}^{\infty} \sum_{p=1}^{\infty}\frac{k_1^2}{p}\exp\left(-\frac{\pi k_1 p\hbar c}{k_BT a}
 \right)\\
 =&-\frac{k_BT \zeta_R(3)}{8 \pi a^3}+\frac{\sqrt{2}(k_BT)^{\frac{5}{2}}}{a^\frac{3}{2}(\hbar c)^{\frac{3}{2}}} \sum_{k_1=1}^{\infty} \sum_{p=1}^{\infty}
\frac{p^{\frac{3}{2}}}{k_1^{\frac{3}{2}} }  K_{\frac{3}{2}}\left(\frac{4\pi k_1p k_B Ta}{\hbar c}
 \right)
-\frac{4\sqrt{2}\pi (k_BT)^{\frac{7}{2}}}{ a^\frac{1}{2}(\hbar c)^{\frac{5}{2}}} \sum_{k_1=1}^{\infty} \sum_{p=1}^{\infty}
\frac{p^{\frac{5}{2}}}{k_1^{\frac{1}{2}}}  K_{\frac{5}{2}}\left(\frac{4\pi k_1p k_B Ta}{\hbar c}
 \right);
\end{split}\end{equation}
and the correction term due to the presence of extra dimensions is
\begin{equation}\label{eq12_22_2}
\begin{split}
&P_{\text{Cas}}^{\parallel, C}(a;T)=  \frac{\hbar c}{a^4}\Biggl\{ -\frac{3 }{8\pi^2   }\sum_{k_1=1}^{\infty}\sum_{\boldsymbol{l}\in \Z^n\setminus\{\mathbf{0}\}} \frac{\sum_{j=1}^{n}\left(\frac{a}{R_j}l_j\right)^2}{k_1^2}    K_2\left(2k_1 \sqrt{\sum_{j=1}^{n}\left(\frac{a}{R_j}l_j\right)^2}\right)
-\frac{1}{4\pi^2   }\sum_{k_1=1}^{\infty}\sum_{\boldsymbol{l}\in \Z^n\setminus\{\mathbf{0}\}} \frac{\left(\sum_{j=1}^{n}\left(\frac{a}{R_j}l_j\right)^2\right)^{\frac{3}{2}}}{k_1}\\&\times K_1\left(2k_1 \sqrt{\sum_{j=1}^{n}\left(\frac{a}{R_j}l_j\right)^2}\right)
 +\frac{\pi k_BTa}{2\hbar c}\sum_{k_1=1}^{\infty}\sum_{\boldsymbol{l}\in\Z^n\setminus\{\mathbf{0}\}}\sum_{p=1}^{\infty}\frac{k_1^2}{p}\exp\left(-\frac{\pi p\hbar c}{k_BT a}
\sqrt{ \frac{1}{\pi^2}\sum_{j=1}^n \left(\frac{a}{R_j}l_j\right)^2+ k_1 ^2 }\right)
 \\&  -\frac{(k_BTa)^2}{2\pi^2(\hbar c)^2}\sum_{\boldsymbol{l}\in \Z^n\setminus\{\mathbf{0}\}}\sum_{p=1}^{\infty} \frac{\sum_{j=1}^n \left(\frac{a}{R_j}l_j\right)^2}{p^2}K_2\left(\frac{p \hbar c}{k_BTa}\sqrt{\sum_{j=1}^{n}\left(\frac{a}{R_j}l_j\right)^2}\right)\Biggr\}\\=&
\frac{\hbar c}{a^4}\Biggl\{\frac{k_B Ta}{4\hbar c} \sum_{k_1=1}^{\infty} \sum_{\boldsymbol{l}\in \Z^n\setminus\{\mathbf{0}\}}\sum_{p=-\infty}^{\infty}\left(
\frac{\frac{1}{\pi^2}\sum_{j=1}^n \left(\frac{a}{R_j}l_j\right)^2+\left(\frac{2pk_BTa}{\hbar c}\right)^2}{k_1^2 }\right)^{\frac{3}{4}}   K_{\frac{3}{2}}\left(2\pi k_1
\sqrt{ \frac{1}{\pi^2}\sum_{j=1}^n \left(\frac{a}{R_j}l_j\right)^2+\left(\frac{2pk_BTa}{\hbar c}\right)^2}\right)
\\&-\frac{\pi k_BTa}{2 \hbar c } \sum_{k_1=1}^{\infty} \sum_{\boldsymbol{l}\in \Z^n\setminus\{\mathbf{0}\}}\sum_{p=-\infty}^{\infty}
\frac{\left(\frac{1}{\pi^2}\sum_{j=1}^n \left(\frac{a}{R_j}l_j\right)^2+\left(\frac{2p k_BTa}{\hbar c}\right)^2\right)^{\frac{5}{4}}}{k_1^{\frac{1}{2}}}  K_{\frac{5}{2}}\left(2\pi k_1
\sqrt{ \frac{1}{\pi^2}\sum_{j=1}^n \left(\frac{a}{R_j}l_j\right)^2+\left(\frac{2p k_BTa}{\hbar c}\right)^2 }\right)\Biggr\}.
\end{split}
\end{equation} \end{widetext}Here $K_{\nu}(z)$ is the modified Bessel function of the second kind. As in the case of a finite piston, the correction term to the Casimir pressure $P_{\text{Cas}}^{\parallel, C}(a;T)$ goes to zero as the size of the extra dimensions goes to zero.

The first and second expressions in \eqref{eq12_22_1} and \eqref{eq12_22_2} correspond respectively to the low and high temperature expansions of the Casimir pressure. Letting $T\rightarrow 0$ in the first expressions, we find that at zero temperature, the Casimir pressure is equal to
\begin{equation}  \label{eq12_22_3}\begin{split}
P_{\text{Cas}}^{\parallel, T=0}(a)=& -\frac{\pi^2 \hbar c  }{480 a^4 }-\frac{3\hbar c }{8\pi^2 a^4  }\sum_{k_1=1}^{\infty}\sum_{\boldsymbol{l}\in \Z^n\setminus\{\mathbf{0}\}}\frac{\sum_{j=1}^{n}\left(\frac{a}{R_j}l_j\right)^2}{k_1^2}  K_2\left(2k_1 \sqrt{\sum_{j=1}^{n}\left(\frac{a}{R_j}l_j\right)^2}\right)
\\&-\frac{\hbar c}{4\pi^2 a^4   }\sum_{k_1=1}^{\infty}\sum_{\boldsymbol{l}\in \Z^n\setminus\{\mathbf{0}\}} \frac{\left(\sum_{j=1}^{n}\left(\frac{a}{R_j}l_j\right)^2\right)^{\frac{3}{2}}}{k_1}  K_1\left(2k_1 \sqrt{\sum_{j=1}^{n}\left(\frac{a}{R_j}l_j\right)^2}\right),\end{split}
 \end{equation}  in agreement with the result of \cite{38}.
The leading term of the temperature correction is equal to $- \pi^2(k_BT)^4/[90(\hbar c)^3]$,   independent of the extra dimensions. The other terms of the thermal correction go to zero exponentially fast when the temperature goes to zero. In the high temperature regime, we read from the second expressions in \eqref{eq12_22_1} and \eqref{eq12_22_2} that the high temperature leading term of the Casimir pressure is linear in temperature:
\begin{equation*}
\begin{split}
P_{\text{Cas}}^{\parallel, T\gg 1}(a)\sim &-k_BT\Biggl\{\frac{  \zeta_R(3)}{8 \pi a^3}+\frac{ 1}{4a^3} \sum_{k_1=1}^{\infty} \sum_{\boldsymbol{l}\in \Z^n\setminus\{\mathbf{0}\}}   \left(
\frac{\frac{1}{\pi^2}\sum_{j=1}^n \left(\frac{a}{R_j}l_j\right)^2 }{k_1^2 }\right)^{\frac{3}{4}}  K_{\frac{3}{2}}\left(2\pi k_1
\sqrt{ \frac{1}{\pi^2}\sum_{j=1}^n \left(\frac{a}{R_j}l_j\right)^2 }\right)\\&-\frac{\pi  }{2 a^3} \sum_{k_1=1}^{\infty} \sum_{\boldsymbol{l}\in \Z^n\setminus\{\mathbf{0}\}}
\frac{\left(\frac{1}{\pi^2}\sum_{j=1}^n \left(\frac{a}{R_j}l_j\right)^2 \right)^{\frac{5}{4}}}{k_1^{\frac{1}{2}}}   K_{\frac{5}{2}}\left(2\pi k_1
\sqrt{ \frac{1}{\pi^2}\sum_{j=1}^n \left(\frac{a}{R_j}l_j\right)^2  }\right)\Biggr\}.\end{split}
\end{equation*}

Notice that \eqref{eq12_22_3}  shows clearly that the zero temperature Casimir pressure is attractive, but \eqref{eq12_22_5}, \eqref{eq12_22_1} and \eqref{eq12_22_2} does not show manifestly that the Casimir pressure at finite temperature is attractive. However, since we have shown that the finite temperature Casimir force \eqref{eq12_18_1} is attractive when the parallel plates are confined in a finite rectangular box, and the Casimir pressure on a pair of infinite parallel plates is obtained by taking the limit   when the transverse dimensions $L_2, L_3$ of the rectangular box approach infinity, we can conclude that the Casimir pressure acting on a pair of infinite parallel plates is attractive at any temperature,  in the presence of any number of extra dimensions. Our results counter the conclusion of \cite{36} which asserted that the Casimir pressure can become repulsive when the temperature is high enough. As pointed out in \cite{38}, the piston analysis   should be the correct way to calculate the Casimir effect on a pair of parallel plates in the presence of extra dimensions, we believe that the results we obtain here is more convincing.

\begin{figure}\centering \epsfxsize=.5\linewidth
\epsffile{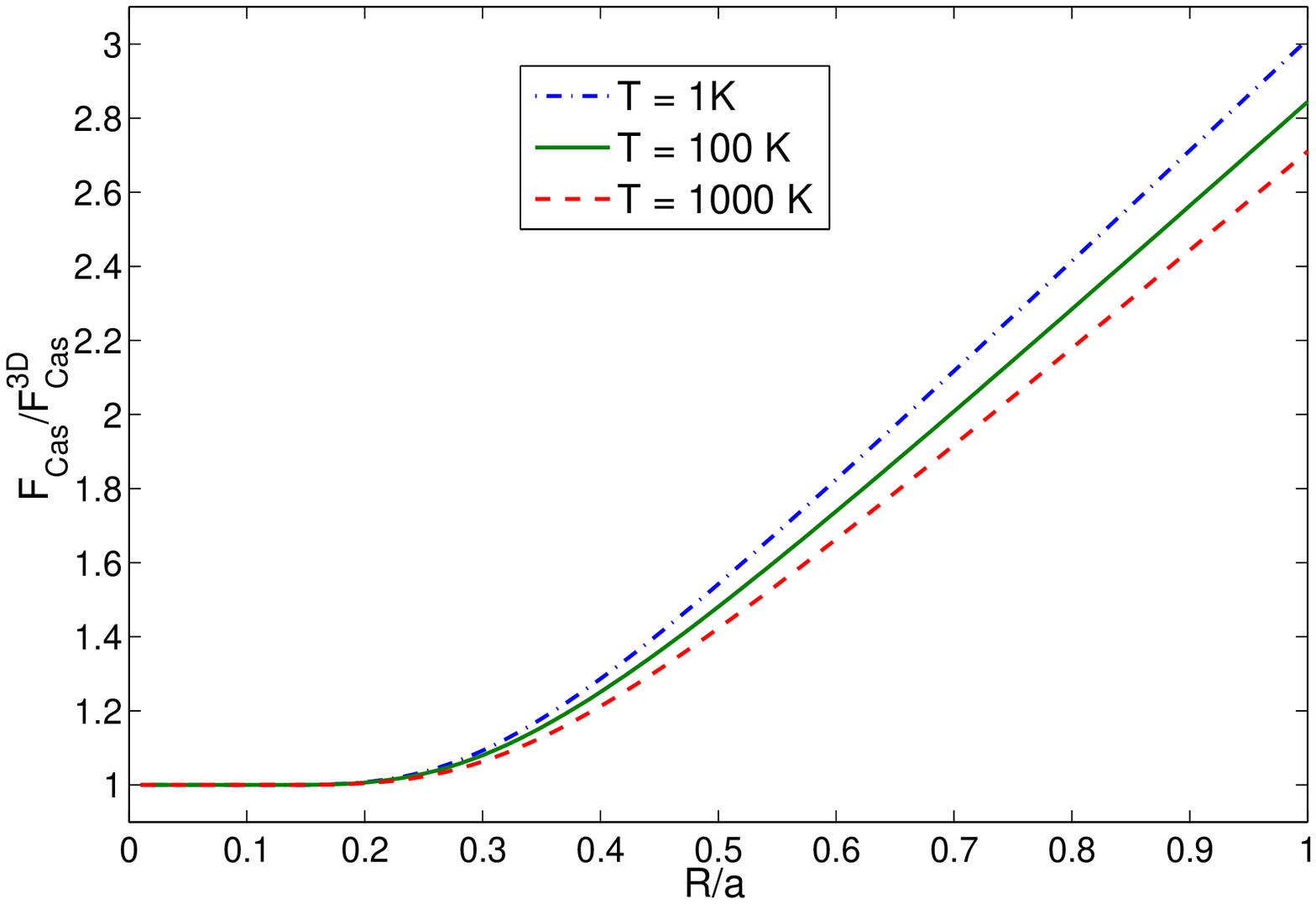}\epsfxsize=.5\linewidth
\epsffile{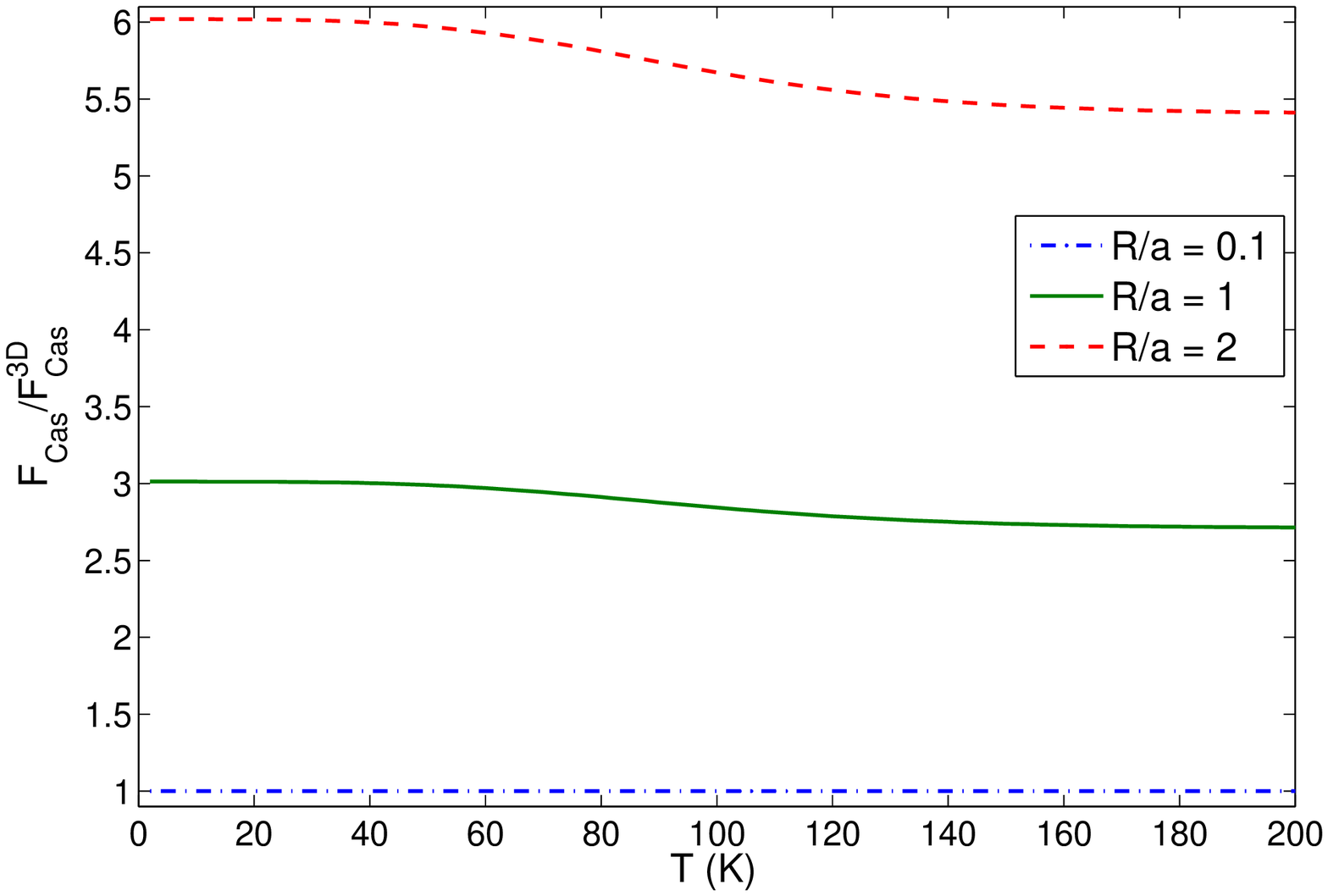}\caption{\label{f2} These graphs show the comparison of the Casimir force acting on a pair of infinite parallel plates separated by a distance $a=10\mu$m in the presence and absence of an extra compactified dimension with radius $R$. }\end{figure}

In FIG. \ref{f2}, we plot the ratio of the Casimir pressure $P_{\text{Cas}}^{\parallel}(a,T)$ acting on a pair of infinite parallel plates in the presence of one extra compactified dimension to the Casimir pressure $P_{\text{Cas}}^{\parallel, 3D}(a,T)$ in the absence of extra dimension. The graphs verify that the Casimir force increases as the size of the extra dimensions increases. They suggest that  the ratio between the Casimir pressures grows linearly with the ratio of  the size of the extra dimension $R$ to the separation between the parallel plates $a$, when the later ratio exceeds a threshold value. On the other hand, they also suggest that the increase of temperature decrease the ratio between the pressures. We shall investigates these aspects analytically in a future work. Notice that when temperature is high enough, the ratio of the Casimir pressures are not much affected by the temperature. This is expected since we have shown that the high temperature leading term of the Casimir pressures are linear in $T$.

Finally we briefly discuss   the general case where the extra dimensions are compactified to a general $n$-dimensional compact connected manifold or orbifold $N^n$ without boundary.   The connected assumption on $N^n$ implies that   the Laplace operator on $N^n$ has exactly one zero eigenvalue corresponding to the constant functions. Let $0=\omega_{N,0}^2/c^2<\omega_{N, 1}^2/c^2\leq \omega_{N, 2}^2/c^2\leq \ldots$ be the complete set of eigenvalues of the Laplace operator on $N^n$ (repeated according to multiplicities). It can then be shown that the Casimir force acting on the piston is given by the same expression \eqref{eq12_18_1} but with summation over $\boldsymbol{l}\in \Z^n$ replaced by summation over $l=0, 1, 2, \ldots$ and $\Lambda_{k_2, k_3, \boldsymbol{l}, p}$ replaced by
\begin{equation*}
\begin{split}
\Lambda_{k_2, k_3, l, p}=\sqrt{\left(\frac{k_2}{L_2}\right)^2+\left(\frac{k_3}{L_3}\right)^2+\frac{\omega_{N,l}^2}{\pi^2}+\left(\frac{2p k_BT}{\hbar c}\right)^2}
\end{split}
\end{equation*}In the limit of infinite parallel plates, it can be shown that the Casimir pressure is given by the expressions analogous to \eqref{eq12_22_1} and \eqref{eq12_22_2}, with the summation over $\boldsymbol{l}\in \Z^n\setminus\{\mathbf{0}\}$ replaced by summation over $l=1, 2, 3, \ldots$ and with the term $\sum_{j=1}^n (l_j/R_j)^2$ replaced by $\omega_{N,l}^2$.  This shows that our conclusion about the attractiveness of the finite temperature Casimir force remains valid for extra dimensions of any geometry.   Since the scaling of the manifold $N^n$ to $RN^n$ will scale the eigenvalues of Laplace operators $\omega_{N,l}^2/c^2$ to $\omega_{N,l}^2/c^2R^2$, the same argument given for the case $N^n=T^n$ shows that the Casimir force is an increasing function of the size $R$ of the extra dimensions. Moreover, in the limit where the extra dimensions vanish, the Casimir force reduces to the corresponding Casimir force in (3+1)-dimensional spacetime. Notice that when $N^n $ is the orbifold $(S^1/\Z^2)^n$, the complete set of eigenvalues of the Laplace operators are given by $\sum_{j=1}^n (l_j/R_j)^2$, $l_j =0, 1, 2, \ldots$. This is actually the case considered in \cite{35, 36, 37}. The assumption that $N^n$ is connected is essential for the connectedness of the spacetime. However, the assumption that $N^n$ is compact and boundaryless is nonessential as one   only need to substitute $\{\omega_{N, l}\}$ by the sequence of eigenvalues of Laplace operator on $N^n$  with appropriate boundary conditions.

We have discussed the behavior of the finite temperature Casimir force acting on a pair of infinite parallel plates due to massless scalar field with Dirichlet boundary conditions in Kaluza-Klein spacetime model. Using the piston setup, we show that at any temperature, the Casimir force is always attractive and at high enough temperature, the strength of the force grows linearly with temperature. We also show that in the limit where the size of the extra dimensions vanishes, the Casimir force reduces to the Casimir force in Minkowski spacetime exponentially fast.

In this work, we only consider massless scalar fields with Dirichlet boundary conditions. However, all the conclusions hold for electromagnetic fields if we assume that the parallel plates are perfect conductors and are of negligible thickness. This is due to the fact that for electromagnetic fields, one only have to multiply every term in \eqref{eq12_18_1} with a suitable integer factor due to the different polarizations of photons. This point shall be discussed in further detail elsewhere.

\end{document}